\documentclass[10pt,twoside]{article}
\usepackage{pkas2}
\usepackage{graphicx}
\makehead{Yasuo Doi et al.}{AKARI/FIS ALL-SKY SURVEY MAPS}

\def\lesssim{\mathrel{\hbox{\rlap{\hbox{\lower4pt\hbox{$\sim$}}}\hbox{$<$}}}}
\def\gtrsim{\mathrel{\hbox{\rlap{\hbox{\lower4pt\hbox{$\sim$}}}\hbox{$>$}}}}

\begin{document}
\sloppy
\pagenumbering{arabic}
\twocolumn[
\pkastitle{27}{1}{2}{2012}
\begin{center}
{\large \bf {\sf 
AKARI FAR-INFRARED ALL-SKY SURVEY MAPS}}\vskip 0.5cm
{\sc Yasuo Doi$^{1}$}\\
  Shinya Komugi$^{2}$
  Mitsunobu Kawada$^{2}$
  Satoshi Takita$^{2}$
  Ko Arimatsu$^{2}$
  Norio Ikeda$^{2}$
  Daisuke Kato$^{2}$
  Yoshimi Kitamura$^{2}$
  Takao Nakagawa$^{2}$
  Takafumi Ootsubo$^{3}$
  Takahiro Morishima$^{3}$
  Makoto Hattori$^{3}$
  Masahiro Tanaka$^{4}$
  Glenn J. White$^{5,6}$
  Mireya Etxaluze$^{5,6,7}$
  and
  Hiroshi Shibai$^{8}$\\
1. Department of Earth Science and Astronomy, the University of Tokyo, Tokyo 153-8902, Japan,
2. Institute of Space and Astronautical Science, Japan Aerospace Exploration Agency, Kanagawa 252-5210, Japan
3. Astronomical Institute, Tohoku University, Miyagi 980-8578, Japan,
4. Center for Computational Sciences, University of Tsukuba, Ibaraki 305-8577, Japan,
5. Department of Physical Sciences, The Open University, Milton Keynes MK7 6AA, England,
6. RALSpace, The Rutherford Appleton Laboratory, Chilton, Didcot, Oxfordshire OX11 0NL, England,
7. Centro de Astrobiolog\'ia (CSIC/INTA), Instituto Nacional de T\'ecnica Aeroespacial, Madrid Spain,
8. Department of Earth and Space Science, Osaka University, Osaka 560-0043, Japan\\
{\it {E-mail: doi@ea.c.u-tokyo.ac.jp} }\\
\normalsize{\it (Received July 1, 2012; Accepted ????)}
\end{center}
\newabstract{
Far-infrared observations provide crucial data for the investigation and characterisation of the properties of dusty material in the Interstellar Medium (ISM), since most of its energy is emitted between $\sim$ 100 and 200 $\mu$m.
We present the first all-sky image from a sensitive all-sky survey using the Japanese {\it AKARI} satellite, in the wavelength range 50 -- 180 $\mu$m.
Covering $>99$\% of the sky in four photometric bands with four filters centred at 65 $\mu$m, 90 $\mu$m, 140 $\mu$m, and 160 $\mu$m wavelengths, this achieved spatial resolutions from 1 to 2 arcmin and a detection limit of $<10$ MJy sr$^{-1}$, with absolute and relative photometric accuracies of $<20$\%. 
\\
All-sky images of the Galactic dust continuum emission enable astronomers to map the large-scale distribution of the diffuse ISM cirrus, to study its thermal dust temperature, emissivity and column density, and to measure the interaction of the Galactic radiation field and embedded objects with the surrounding ISM.
In addition to the point source population of stars, protostars, star-forming regions, and galaxies, the high Galactic latitude sky is shown to be covered with a diffuse filamentary-web of dusty emission that traces the potential sites of high latitude star formation.
We show that the temperature of dust particles in thermal equilibrium with the ambient interstellar radiation field can be estimated by using 90 $\mu$m, 140 $\mu$m, and 160 $\mu$m data.
The FIR {\it AKARI} full-sky maps provide a rich new data set within which astronomers can investigate the distribution of interstellar matter throughout our Galaxy, and beyond.
\\
\vskip 0.5cm
{\em key words:} Surveys -- Atlases -- ISM: general -- Galaxy: general -- Galaxies: general -- Infrared: ISM -- Infrared: galaxies}
\vskip 0.15cm  \flushbottom
]

\newsection{INTRODUCTION}
Infrared continuum emission is ubiquitously observed across the celestial sky and is attributed to the thermal emission from interstellar dust particles.
The spectral energy distribution (SED) of the dust continuum emission has been observed for various astronomical objects including diffuse interstellar medium, star-forming regions and galaxies.
The dust continuum SED has its peak at far-infrared (30--300 $\mu$m; FIR) wavelengths, with approximately two thirds of the energy being radiated at $\lambda \geq 50\ \mu$m (Draine 2003; also see Compi{\`e}gne et al. 2011).
Since the stellar radiation is mainly absorbed by large grains (LGs) in interstellar space, and converted to FIR continuum emission, it is important to measure the total FIR continuum emission energy as it is a good tracer of total stellar radiation energy and thus a good indicator of the star-formation activity (Kennicutt 1998).
The $\beta = 2$ modified black-body spectrum of temperature 15 -- 30 K has its peak at 100 -- 200 $\mu$m.
It is thus essential to observe the FIR continuum, including this wavelength range, to measure the total FIR emission energy from LGs to trace the star-formation activity.

An all-sky survey of FIR continuum was pioneered by the IRAS satellite (Neugebauer et al. 1984), which was, until the completion of the {\it AKARI} mission, the only available FIR all-sky survey with high spatial resolution.
Given the importance of the FIR photometric data, it is valuable to upgrade the data by improving spatial resolution, and by adding more photometric bands, especially at longer wavelengths.

For that purpose, we performed new all-sky survey observations with {\it AKARI} (Murakami et al. 2007).
%
In this paper, we describe the quality of the image data and its capability to investigate spatial distribution of ISM and its SED with improved spatial resolutions for the whole sky.

\newsection{Observation and Data Reduction}

The {\it AKARI} satellite (Murakami et al. 2007) was launched in February 2006 and the all-sky survey was performed between March 2006 -- August 2007, which was the cold operational phase of the satellite with liquid Helium cryogen.
The all-sky survey observations in FIR wavebands were carried out by using an FIR instrument called the Far-Infrared Surveyor (FIS: Kawada et al. 2007), which covered the 50 -- 180 $\mu$m waveband (Kawada et al. 2007).

Four photometric bands are used for the all-sky survey.
Two of four bands had a broader wavelength coverage and continuously covered the whole waveband range: the WIDE-S band (50--110 $\mu$m, centred at 90 $\mu$m) and the WIDE-L band (110--180 $\mu$m, centred at 140 $\mu$m).
The other two bands had narrower wavelength coverage and measured both the shorter and the longer ends of the wavebands: the N60 band (50--80 $\mu$m, centred at 65 $\mu$m) and the N160 band (140--180 $\mu$m, centred at 160 $\mu$m).

The observed fraction of the sky during the survey period (the cold operational phase) is tabulated in Table 1.
About 97 \% of the sky was multiply surveyed in the four photometric bands.
\begin{table}[!t]
  \begin{center}
    \scriptsize
    \bf{\sc  Table 1.}\\
    \sc{Scan coverage of the survey observations} \\
    \label{tab:spec}
    \begin{tabular}{@{}c@{}ccccc@{}}
      \\ \hline \hline

      Spatial scans~ & Single & Multiple & $\geq 5$ & $\geq 10$ & Unscanned\\
      \hline
      N60\dotfill & 99.1\% & 96.9\% & 60.3\% & 13.5\% & 0.9\% \\ 
      Wide-S\dotfill & 99.1\% & 97.0\% & 61.0\% & 13.8\% & 0.9\% \\ 
      Wide-L\dotfill & 99.5\% & 98.4\% & 78.7\% & 25.9\% & 0.5\% \\ 
      N160\dotfill & 99.5\% & 98.4\% & 76.8\% & 24.1\% & 0.5\% \\
      \hline
      4 bands\dotfill & 99.1\% & 96.8\% & 60.0\% & 13.2\% & 0.9\% \\
      \hline
      \end{tabular}
  \end{center}
\end{table}

The data obtained during the observations was pre-processed using the {\it AKARI} pipeline tool, which was originally optimised for point source extraction (Yamamura et al. 2009).
Following this initial pre-processing, a separate pipeline was then used to recover the large-scale spatial structures accurately.
The main properties of the diffuse mapping pipeline are: 1) transient response correction of the detector signal, 2) subtraction of zodiacal emission, 3) image processing, 4) image destriping, and 5) recursive deglitching and flat-fielding of processed image.
Processes 1) and 2) were performed on the time-series data and 3) -- 5) were performed on the image plane data.
The details of these procedures are found in Doi et al. (2009) and Doi et al. (2012).

\newsection{Results}

\begin{figure}[!ht]
  \resizebox{\hsize}{!}{\includegraphics{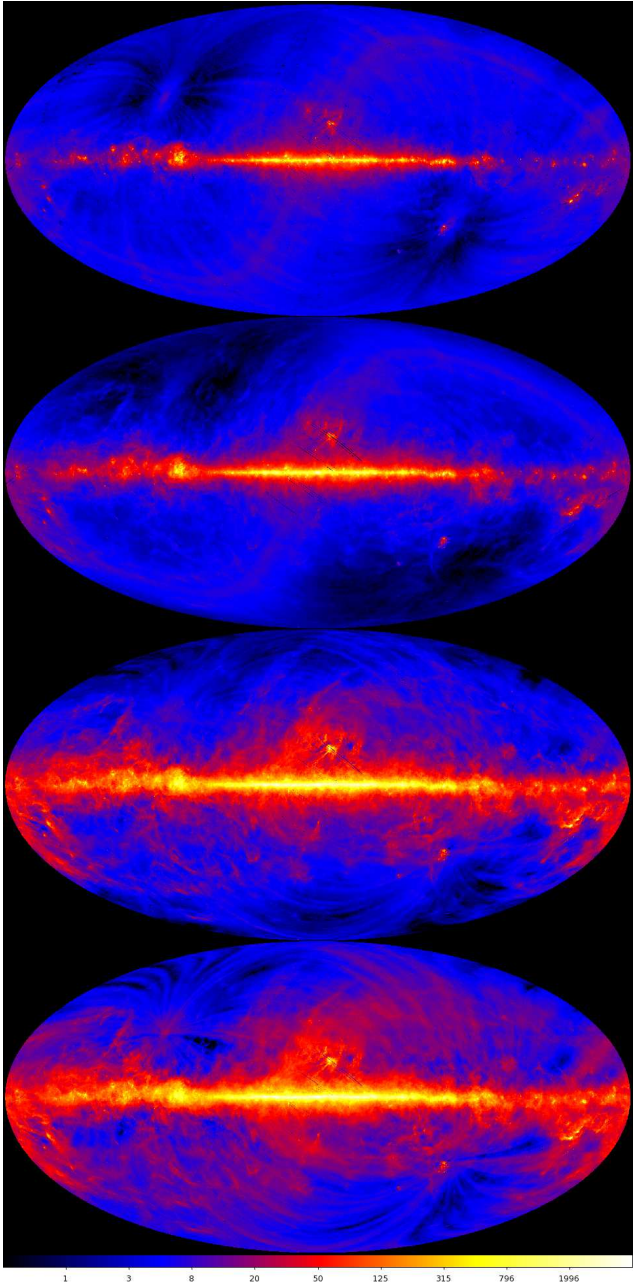}}
  \caption{All-sky FIR intensity maps observed in the {\it AKARI} diffuse survey. Four bands are displayed from top to bottom: N60 (65 $\mu$m), WIDE-S (90 $\mu$m), WIDE-L (140 $\mu$m), and N160 (160 $\mu$m). The intensity units are in MJy sr$^{-1}$.}
  \label{fig:4plates}
\end{figure}

Intensity maps from the all-sky survey in the four photometric bands are shown in Fig.~\ref{fig:4plates}.
Detailed photometry with high spatial resolution has been achieved for the whole sky.
The overall spatial distribution shows a clear wavelength dependence.
The intensity distributions of the emission in shorter wavelengths (65 $\mu$m \& 90 $\mu$m) show the concentrated distribution on the Galactic plane, indicating a tight correlation with star-formation activity by tracing high ISRF regions at and around star-formation regions.
On the other hand, relatively intense diffuse emission is found in emission at longer wavelengths (140 $\mu$m \& 160 $\mu$m), especially from the regions at high Galactic latitudes.
Emission from lower temperature dust in high latitude cirrus clouds is thus well traced by longer wavelength observations and the ISM total mass distribution is well traced by these observations.

Zodiacal emission has been subtracted from the images and the residual emission can be noticed in shorter wavelength images.
The residual emission is attributed to the Zodiacal dust bands, as we only subtract a spatially smooth component of the Zodiacal emission and do not subtract the emission with smaller scale structures.
Taking advantage of the higher spatial resolutions of the {\it AKARI} full-sky images, we will investigate these structures down to arc-minute spatial scales to model and discriminate them from other astronomical emissions (Ootsubo et al. 2012).

We have primarily calibrated the data using the more sensitive slow-scan data (Shirahata et al. 2009), as the spatial scan speeds are slow enough so that the data have better sensitivity and are virtually not affected by the transient response of the detector.
Komugi et al. (2012) estimate the 1-$\sigma$ detection limit of the survey data as 12 MJy sr$^{-1}$, 2 MJy sr$^{-1}$, 7 MJy sr$^{-1}$, and 4 MJy sr$^{-1}$ for N60, WIDE-S, WIDE-L, and N160 bands, respectively.
The efficacy of the linearity correction of the detector signal was also checked with slow scan images that observed the same position in the sky and is not affected by the detector transient response due to its longer integration time.
We confirm that a good linearity correction has been achieved for all the bands, from the faintest detector signals up to the bright signals around the Galactic center of $>10$ [GJy sr$^{-1}$], with relative accuracy of $\sim 20\%$.
It is important to note that the good linearity correction achieved for the data also confirms the successful correction of the spatial distortion of the data caused by the transient response of the detector.
The linearity and absolute intensity have been checked with other IR observations by ISOPHOT (Kessler et al. 1996), IRAS/IRIS (Miville-Desch{\^e}nes \& Lagache 2005), and COBE/DIRBE (Hauser et al. 1998).
Takita et al. (2012) concluded that the correlation of {\it AKARI} survey data with other FIR data confirms the reliable calibration.

\newsection{Discussion}

The aim of our survey observation is twofold, 1) to obtain the global distribution of ISM with higher spatial resolutions, and 2) to study the SEDs and to evaluate the colour temperature of LGs with multi-band photometry that has good wavelength coverage.
In the following discussion, we investigate the characteristics of the survey data from these two viewpoints.

\newsubsection{Spatial Power Spectra}

One of the key characteristics of the {\it AKARI} FIR survey is its large dynamic range that enables simultaneous recovery of both large scale spatial structures and more localised small scale variations of the FIR emission.
This characteristic can be examined by calculating the spatial power spectra of the cirrus distribution taken from the {\it AKARI} diffuse all sky survey, as the cirrus power spectra can be well represented by power-law spectra (see e.g., Miville-Desch{\^e}nes et al. 2010; Martin et al. 2010).

Miville-Desch{\^e}nes et al. (2010) derived the IRAS/IRIS 100 $\mu$m and Herschel/SPIRE 250 $\mu$m combined spatial power spectrum at the Polaris flare region and confirmed that the power-law nature of the cirrus power spectrum is kept down to sub-arcminute scales, with a power-law index $\gamma = -2.65 \pm 0.10$ on scales $0.025 < k < 2 {\rm\ [arcmin^{-1}]}$ (also see Martin et al. 2010).
Since the Polaris flare is a high Galactic latitude cirrus cloud with virtually no sign of star-formation activity (Martin et al. 2010), the region is well suited to assess the nature of the cirrus power spectrum.

\begin{figure}[!ht]
  \resizebox{\hsize}{!}{\includegraphics{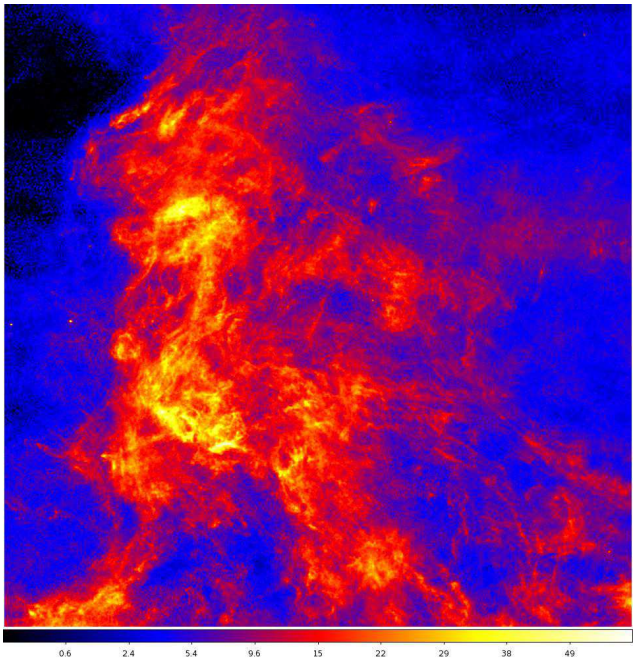}}
  \caption{WIDE-L 140$\mu$m intensity image of the Polaris Flare region in Galactic coordinates centering at $l = 121^{\circ}$ and $b = 28^{\circ}$. The size of the images are $15^{\circ} \times 15^{\circ}$.
The intensity units are in MJy sr$^{-1}$.}
  \label{fig:PolarisFlare}
\end{figure}

\begin{figure}[!ht]
  \resizebox{\hsize}{!}{\includegraphics{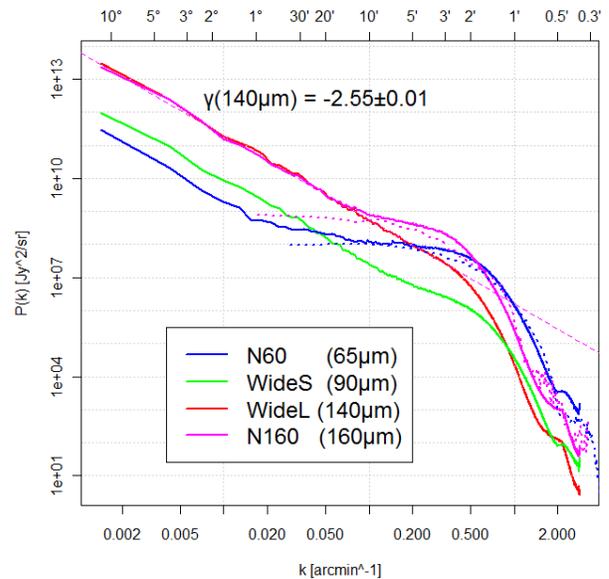}}
  \caption{Spatial power spectra of {\it AKARI} intensities observed in the Polaris flare region shown in Fig.~\ref{fig:PolarisFlare} (solid lines). Dotted lines show point spread functions (PSFs) of N60 and N160 bands. Higher spatial frequency components of N60 \& N160 power spectra are dominated by high frequency noise and well represented by their PSFs.}
  \label{fig:PolarisFlare_powerspectra}
\end{figure}

The {\it AKARI} FIR image of the Polaris flare region and its spatial power spectra are shown in Figs.~\ref{fig:PolarisFlare} \& \ref{fig:PolarisFlare_powerspectra}.
The cirrus spectrum can be represented by a single power-law, as indicated by many other authors (Miville-Desch{\^e}nes et al. 2010 and the references therein).
The power law indices are identical for the four wavebands.
It is estimated as $\gamma_{140\mu m} = 2.55 \pm 0.01$ by fitting the 140 $\mu$m spectrum in $k = 0.01 $--$ 0.33$ [arcmin$^{-1}$] wavenumber range, and the estimated $\gamma$ is consistent with that by Miville-Desch{\^e}nes et al. (2010) within the range of error.
The spectra clearly show the wide dynamic range of our observation, as the spatial power component is well retrieved from $> 10^{\circ}$ large scale distribution to $\sim 3'$ small scale structures at 140 $\mu$m and even smaller structures at 90 $\mu$m.

On the other hand, excess components in the power spectrum are found at the high spatial frequencies in the two narrow bands (N60 \& N160).
This is due to the higher noise levels of these two bands compared to the wide bands (WIDE-S \& WIDE-L).
The spatial dynamic range of the narrow bands is thus limited for the faint cirrus component.
Better recovery of these high frequency spatial component of fainter regions could be obtained by adjusting the correction function of the detector transient response, as the correction is primarily an amplification of the high spatial frequency component and thus causes excess noise at fainter regions (see Doi et al. 2009).
This adjustment will be considered in a future update of the image data.

\newsubsection{Spectral Energy Distribution}

\begin{figure}[!ht]
  \resizebox{\hsize}{!}{\includegraphics{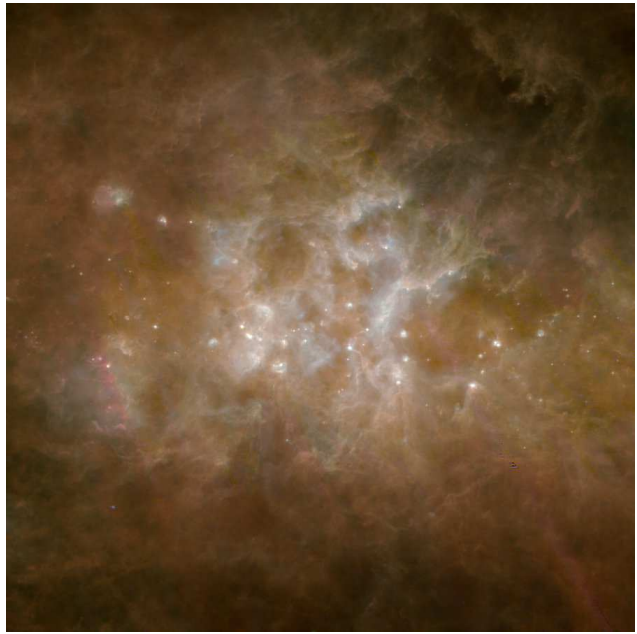}}
  \caption{An intensity map the Cyngus X region in Galactic coordinates centering at $l = 80^{\circ}$ and $b = +1^{\circ}$ with the image size of $15^{\circ} \times 15^{\circ}$. The image is a three-colour composite of N60 (blue), WIDE-S (green), and WIDE-L (red).}
  \label{fig:CygX}
\end{figure}

\begin{figure}[!ht]
  \resizebox{\hsize}{!}{\includegraphics{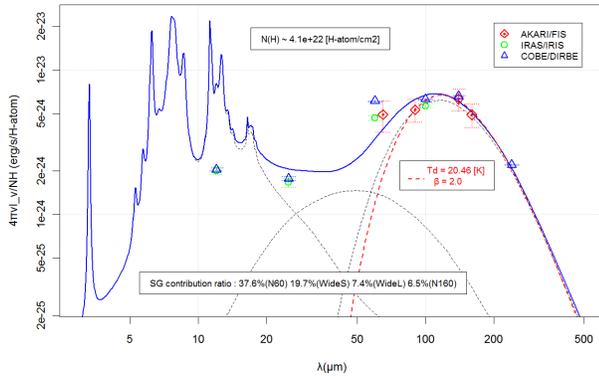}}
  \caption[]{Spectrum Energy Distribution at Cygnus X region ($l=80^{\circ}$, $b=+1^{\circ}$) in units of $4\pi\nu I_{\nu} {\rm N_H}$ ergs s$^{-1}$ H-atom$^{-1}$.
The {\it AKARI} FIR data as well as IRAS/IRIS (Miville-Desch{\^e}nes \& Lagache 2005) and COBE/DIRBE (Hauser et al. 1998) data are shown.
The data are averaged over the 0.7$^{\circ}$ DIRBE beam, and mean values and those standard deviations are indicated.
The red dotted line is an estimation of the LGs colour temperature by fitting 90$\mu$m, 140$\mu$m, 160$\mu$m data with a single modified blackbody spectrum.
The blue solid line is a model fit of the four {\it AKARI} data by DustEM SED model (Compi{\`e}gne et al. 2011). The decomposition of the model emission into three dust components (PAHs, SGs and LGs) are indicated as black dotted lines.
Relative contribution of emissions from SGs to each {\it AKARI} intensity is also indicated in the figure.
}
  \label{fig:CygX_SED}
\end{figure}

Another key aspect of the {\it AKARI} FIR survey is its resolution of the spectral peak of the dust SED with four wavebands.
Figs.~\ref{fig:CygX} \& \ref{fig:CygX_SED} show the intensity distribution of the Cygnus X region and its SED at the centre of the region ($l=80^{\circ}$, $b=+1^{\circ}$).
Ancillary data at the same position by IRAS/IRIS (Miville-Desch{\^e}nes \& Lagache 2005) and COBE/DIRBE (Hauser et al. 1998) and their spatial variation within the $0.7^{\circ}$ DIRBE beam are also shown.
The {\it AKARI} data have the largest spatial variation because of higher spatial resolutions.

Good coincidence between {\it AKARI} and the ancillary data is found.
The intensity at 140$\mu$m is the highest, confirming that the peak of dust SED is well covered by the {\it AKARI} data.
WIDE-S, WIDE-L, N160 and ancillary data at $\geq 90 \mu$m are well fit with a modified blackbody of T$=20.5$ [K] and $\beta = -2$,
indicating that the {\it AKARI} FIR data can be a good tracer of the temperature and the total amount of LGs.
Since {\it AKARI} has higher spatial resolution comparing to other all-sky survey observations as well as huge spatial dynamic range at this wavelength range, we can make a detailed investigation of the spatial distribution of temperature, and the total amount of LGs.

On the other hand, the N60 observations, and other data at $60 \mu$m show significant excess from a modified blackbody, indicating a significant contribution from either higher temperature LGs along the line of sight or stochastically heated small grains (SGs).
A fit using the DustEM SED model by Compi{\`e}gne et al. (2011) with an assumption of uniform incident radiation is also shown in Fig.~\ref{fig:CygX_SED}.
A significant contribution to the ratio of emission from SGs to the total observed intensity is estimated from the model, not only for N60 (38\%), but also for WIDE-S (20\%).
Hence it is important to note that the $90\mu {\rm m}/140\mu {\rm m}$ colour temperature could overestimate the dust temperature, though the ambiguity is much smaller than the widely referred to issues with the $60\mu {\rm m}/100\mu {\rm m}$ IRAS colour temperature.

The shorter wavelength excess emission is, however, not completely explained by the modelled emission from small dust particles with uniform irradiation.
Temperature variation of dust particles, or intensity variations of interstellar radiation field, along the line of sight also should be taken into account.
The wealth of spatial structures and colour variations shown in Fig.~\ref{fig:CygX} clearly indicate the complexity of the ISM and its irradiation environment.
The newly achieved {\it AKARI} FIR high-spatial resolution images of the whole sky should be a powerful tool to investigate the detailed spatial structure of ISM and its physical environment.

\newsection{Conclusion}

We have processed a full sky image of the FIR {\it AKARI} survey at $65 \mu$m, $90 \mu$m, $140 \mu$m and $160 \mu$m.
The images achieve a detection limit of $<10$ MJy sr$^{-1}$ with absolute and relative photometric accuracies of $<20$\%.
Together with the $>99$\% coverage of the whole sky, the high spatial resolution of the FIR {\it AKARI} survey from 1 to 2 arcmin achieves an unprecedented spatial dynamic range of $>10^4$.
A good wavelength coverage from 50$\mu$m to 180$\mu$m with four photometric bands makes the data a good tracer of temperature and the total amount of the dust component in the ISM.
The wealth of spatial structures and SED variations found in the image clearly indicate the complexity of the ISM and its irradiation environment.
The FIR {\it AKARI} image is a new powerful tool to investigate the detailed nature of ISM from small scales to the full sky.

\end{description}


\references
\begin{description}

\bibitem{Beichman, C.~A., Neugebauer, G., Habing, H.~J., Clegg, P.~E., \& Chester, T.~J.\ 1988, Infrared astronomical satellite (IRAS) catalogs and atlases.~Volume 1: Explanatory supplement, 1}
\bibitem{Boggess, N.~W., Mather, J.~C., Weiss, R., et al.\ 1992, The COBE mission - Its design and performance two years after launch, ApJ, 397, 420}
\bibitem{Compi{\`e}gne, M., Verstraete, L., Jones, A., et al.\ 2011, The global dust SED: tracing the nature and evolution of dust with DustEM, A\&AP, 525, A103}
\bibitem{Doi, Y., et al.\ 2009, Far-Infrared All Sky Diffuse Mapping with AKARI, Astronomical Society of the Pacific Conference Series, 418, 387}
\bibitem{Doi, Y., et al. 2012, in preparation.}
\bibitem{Draine, B.~T.\ 2003, Interstellar Dust Grains, ARA\&A, 41, 241}
\bibitem{Hauser et al.\ 1998, COBE Diffuse Infrared Background Experiment (DIRBE) Explanatory Supplement ed. M.G. Hauser, T. Kelsall, D. Leisawitz, and J. Weiland COBE Ref. Pub. No. 97-A (Greenbelt, MD: NASA/GSFC)}
\bibitem{Kawada, M., Baba, H., Barthel, P.~D., et al. 2007, The Far-Infrared Surveyor (FIS) for AKARI, PASJ, 59, S389}
\bibitem{Kennicutt, Jr., R.~C. 1998, Star Formation in Galaxies Along the Hubble Sequence, ARA\&A, 36, 189}
\bibitem{Kessler, M.~F., Steinz, J.~A., Anderegg, M.~E., et al.\ 1996, The Infrared Space Observatory (ISO) mission, A\&AP, 315, L27}
\bibitem{Komugi, S., Doi, Y., et al. 2012, in preparation.}
\bibitem{Martin, P.~G., Miville-Desch{\^e}nes, M.-A., Roy, A., et al.\ 2010, Direct estimate of cirrus noise in Herschel Hi-GAL images, A\&AP, 518, L105}
\bibitem{Miville-Desch{\^e}nes, M.-A., \& Lagache, G.\ 2005, IRIS: A New Generation of IRAS Maps, ApJS, 157, 302}
\bibitem{Miville-Desch{\^e}nes, M.-A., Martin, P.~G., Abergel, A., et al.\ 2010, Herschel-SPIRE observations of the Polaris flare : structure of the diffuse interstellar medium at the sub-parsec scale, A\&AP, 518, L104}
\bibitem{Murakami, H., Baba, H., Barthel, P. D., et al. 2007, The Infrared Astronomical Mission AKARI, PASJ, 59, S369}
\bibitem{Neugebauer, G., Habing, H.~J., van Duinen, R., et al.\ 1984, The Infrared Astronomical Satellite (IRAS) mission, ApJL, 278, L1}
\bibitem{Ootsubo, T., Doi, Y., et al. 2012, in preparation.}
\bibitem{Pilbratt, G.~L., Riedinger, J.~R., Passvogel, T., et al.\ 2010, Herschel Space Observatory. An ESA facility for far-infrared and submillimetre astronomy, A\&AP, 518, L1}
\bibitem{Shirahata, M., Matsuura, S., Hasegawa, S., et al. 2009, Calibration and Performance of the AKARI Far-Infrared Surveyor (FIS) -- Slow-Scan Observation Mode for Point-Sources, PASJ, 61, 737}
\bibitem{Takita, S., Doi, Y., et al. 2012, in preparation.}
\bibitem{Werner, M.~W., Roellig, T.~L., Low, F.~J., et al.\ 2004, The Spitzer Space Telescope Mission, ApJS, 154, 1}
\bibitem{Yamamura, I., et al.\ 2009, Release of the AKARI-FIS Bright Source Catalogue ƒÀ-1, Astronomical Society of the Pacific Conference Series, 418, 3}
\end{document}